\documentclass[runningheads]{llncs}

\usepackage{times}
\usepackage{physics}
\usepackage{amsmath}
\usepackage{amssymb}
\usepackage{xcolor}
\usepackage{acronym}
\usepackage{graphicx}
\usepackage{subcaption} 
\usepackage{hyperref}
\usepackage{tablefootnote}
\usepackage{tabularx}
\usepackage{threeparttable}
\usepackage{booktabs}
\usepackage{multirow}
\usepackage{siunitx}
\usepackage[numbers,sort&compress]{natbib}  
\usepackage[skip=0pt]{caption}
\usepackage[inline]{enumitem}
\usepackage[linesnumbered,lined,ruled,commentsnumbered]{algorithm2e}

\usepackage{hyperref}
\hypersetup{
  colorlinks   = true,    
  urlcolor     = blue,    
  linkcolor    = blue,    
  citecolor    = red      
}

\newcommand{\header}[1]{\vspace*{1mm}\noindent\textbf{#1}}

\acrodef{DS}{diversity score}
\acrodef{NBR}{next basket recommendation}
\acrodef{RNN}{recurrent neural network}
\acrodef{KNN}{$k$-nearest neighbor}
\acrodef{GNN}{graph neural network}
\acrodef{DCTR}{document-based click-through rate model}
\acrodef{PHR}{personalized hit ratio}
\acrodef{NDCG}{normalized discounted cumulative gain}
\acrodef{TREx}{two-step repetition-exploration}
\acrodef{AWRF}{attention-weighted rank fairness}
\acrodef{EUR}{exposed utility ratio}
\acrodef{RUR}{realized utility ratio}

\acrodef{IAA}{inequality of amortized attention}
\acrodef{EEL}{expected exposure loss}
\acrodef{EER}{expected exposure relevance}
\acrodef{EED}{expected exposure disparity}
\acrodef{logDP}{log of demographic parity}
\acrodef{DP}{demographic parity}
\acrodef{logEUR}{exposed utility ratio}

\acrodef{logRUR}{realized utility ratio}
\acrodef{MILP}{mixed-integer linear programming}
\acrodef{ILP}{integer linear programming}
\acrodef{RBP}{rank-biased precision}

\acrodef{MMR}{maximal marginal relevance}

\acrodef{PIF}{personalized item frequency}
\acrodef{RADiv}{repeat-bias-aware diversity optimization}
\acrodef{RAIF}{repeat-bias-aware item fairness optimization}

\newcommand{\negskip}{\vspace*{-1.5mm}}
\newcommand{\halfnegskip}{\vspace*{-0.75mm}}
\author{
Yuanna Liu\orcidID{0000-0002-9868-6578}
\and 
Ming Li\orcidID{0000-0001-7430-4961}
\and 
Mohammad Aliannejadi\orcidID{0000-0002-9447-4172}
\and 
Maarten de Rijke\orcidID{0000-0002-1086-0202}}

\institute{University of Amsterdam\\
\email{\{y.liu8,m.li,m.aliannejadi,m.derijke\}@uva.nl}}

\begin{document}

\title{Repeat-bias-aware Optimization of Beyond-accuracy Metrics for Next Basket Recommendation}
    
\titlerunning{Repeat-bias-aware Optimization for Next Basket Recommendation}
%

\maketitle              

\begin{abstract}
        
In \ac{NBR} a set of items is recommended to users based on their historical basket sequences. 
In many domains, the recommended baskets consist of both repeat items and explore items. 
Some state-of-the-art NBR methods are heavily biased to recommend repeat items so as to maximize utility. 
The evaluation and optimization of beyond-accuracy objectives for NBR, such as item fairness and diversity, has attracted increasing attention.
How can such beyond-accuracy objectives be pursued in the presence of heavy repeat bias?
We find that only optimizing diversity or item fairness without considering repeat bias may cause NBR algorithms to recommend more repeat items. 
To solve this problem, we propose a model-agnostic repeat-bias-aware optimization algorithm to post-process the recommended results obtained from NBR methods with the objective of mitigating repeat bias when optimizing diversity or item fairness. 
We consider multiple variations of our optimization algorithm to cater to multiple NBR methods.
Experiments on three real-world grocery shopping datasets show that the proposed algorithms can effectively improve diversity and item fairness, and mitigate repeat bias at acceptable Recall loss. 

\keywords{Next basket recommendation   \and Repeat bias  \and Beyond-accuracy metrics \and Re-ranking}
\end{abstract}

\acresetall

\section{Introduction}

In \ac{NBR} a recommender system is meant to recommend a set of items at once~\cite{li2023intent}. 
In many e-commerce scenarios in which \ac{NBR} are deployed users display \emph{repetitive} consumption behavior (e.g., purchasing milk every week, or listening to the same song during workouts) as well as \emph{exploratory} consumption behavior (e.g., buying Christmas gifts) at the same time. Hence, items in a recommended basket can be categorized into repeat items and explore items, depending on whether a user has consumed an item before.

\header{Beyond-accuracy metrics and \ac{NBR}.} Many machine learning techniques have been applied to optimize the accuracy achieved on the \ac{NBR} task, based on \acp{KNN}~\cite{hu2020modeling,faggioli2020recency,naumov2023time}, \acp{RNN}~\cite{yu2016dynamic}, or \acp{GNN}~\cite{yu2020predicting}. 
However, measuring and optimizing beyond-accuracy objectives for \ac{NBR} remains largely unexplored. 
Of particular interest are \emph{diversity}, to combat the problem of recommendation homogeneity and generate diversified baskets~\cite{leng2020recurrent,sun2023generative}, and \emph{item fairness}, to ensure a fair distribution of the exposure assigned to different groups of items~\cite{liu2024measuring}. 
Optimizing for beyond-accuracy metrics such as diversity and item fairness in the context of \ac{NBR} is made more complex than in other recommendation scenarios due to the presence of repeat bias.

\header{Repeat bias in NBR.}
Experiments have shown that repeat items contribute most of the accuracy performance~\cite{li2023next, li-2023-repetition, li-2023-repetition-offline, li-2023-will}. 
This is because repetition prediction is an easier task, typically with just dozens or hundreds of candidate items and explicit user feedback~\cite{li2024we}. 
However, explore items hold great potential for long-term value in particular and for beyond-accuracy goals in general. 
E.g., exploring uncertain regions enhances exposure to new and long-tail content, reshaping the overall distribution of the contents, which ultimately improves long-term user experience~\cite{su2024long}. 
Therefore, when curating sets that include both previously seen and new items, it is important to strike the right balance between the repeat and explore categories~\cite{ariannezhad2023complex}.
Importantly, finding a good balance between repeat and explore items in \ac{NBR} is made more complex when beyond-accuracy metrics are considered~\cite{li2024we}:
if most utility is obtained from a relatively small number of items (the repeat items), there is little room to improve beyond-accuracy metrics without sacrificing utility. 

\header{Repeat-bias-aware optimization for \ac{NBR}.}
For the generic top-$k$ recommendation task, re-ranking is a direct and effective method for improving beyond-accuracy performance, e.g., through greedy algorithms and constrained optimization~\cite{zhao2023fairness}. 
The essence of re-ranking is to determine and exploit an effective trade-off between predicted relevance score and beyond-accuracy metrics.  
However, only optimizing diversity or item fairness may lead to increased repeat bias in some cases according to our experiments, i.e., to an increase in the deviation of the repeat ratio in recommended baskets vs.\ in ground truth baskets~\citep{tran2024transformers}.
Even though previous research highlights the problem of repeat bias, no one has tried to optimize to mitigate it. To the best of our knowledge, we are the first to optimize repeat bias jointly with other metrics to seek a balance among (mitigating) repeat bias, accuracy, and beyond-accuracy metrics.

To solve this optimization problem, we propose a model-agnostic \ac{RADiv} algorithm and a \ac{RAIF} algorithm based on \acfi{MILP} for \ac{NBR}. 
The repeat ratio (or rather: reducing it) is one of the optimization objectives. 
The proposed algorithms optimize for predicted relevance, diversity (or item fairness), and repeat ratio simultaneously. 
We offer several flavors of our \ac{MILP}-based optimization algorithm, to benefit from the peculiarities of different families of \ac{NBR} methods.
To the best of our knowledge, we are the first to apply a re-ranking algorithm to jointly optimize beyond-accuracy metrics and repeat bias in \ac{NBR}. 

Summarizing, the \textbf{main contributions} of the paper are:
\begin{itemize}[leftmargin=*,nosep]
    \item We propose repeat-bias-aware optimization algorithms (\ac{RADiv} and \ac{RAIF}) for \ac{NBR}, which mitigate repeat bias while improving the diversity and item fairness of recommended baskets.

    \item We extend these algorithms to multiple \ac{NBR} paradigms, including ones that merge and optimize items from separate repetition and exploration models.

    \item We conduct experiments on three retail datasets and find that the proposed \ac{RADiv} and \ac{RAIF} algorithms can significantly improve diversity, and item fairness and mitigate repeat bias with an acceptable loss in Recall. 

\end{itemize}

\negskip
\section{Related Work}
\label{related}


\header{Beyond-accuracy objectives in recommender system.}
In recommendation, optimizing only for accuracy measures is limiting and misguided~\cite{bauer2024values}. 
There is a growing interest in beyond-accuracy metrics, which measure other recommendation qualities~\citep{derijke-2023-beyond}. \citet{kaminskas2016diversity} study the definitions and metrics of diversity, serendipity, novelty, and coverage. They implement re-ranking strategies for these beyond-ac\-curacy metrics to investigate correlations between the two objectives.

Fairness of recommender systems has sparked much research in the community, considering multiple stakeholders of typical e-commerce platforms. From the user side, user fairness requires that a fair system should provide the same recommendation quality for different user groups~\cite{li2021user}. From the item side, the goal is to measure the exposure assigned to each item, or each group, and evaluate this distribution to ensure fair principles, such as statistical parity or equal opportunity~\cite{raj2022measuring}. Usually, the exposure of an item in a ranked list is computed based on a user browsing model~\cite{chapelle2009expected,moffat2008rank}.

\header{Beyond-accuracy objectives in \ac{NBR}.}
In \ac{NBR}, most deep-learning-based models predict the top-$k$ relevant items via a user representation to form a basket. This paradigm leads to the problem of over-homogenization of the recommended baskets~\cite{sun2023generative}. To address this problem, \citet{sun2023generative} apply an autoregressive item-level decoder to generate items one by one to ensure diversified baskets. \citet{leng2020recurrent} employ a deconvolutional network to generate diverse \ac{NBR} results. Regarding item fairness, \citet{liu2024measuring} reproduce a set of item fairness metrics to evaluate representative \ac{NBR} methods. \citet{li2024we} propose a framework to identify short-cuts in achieving better accuracy and beyond-accuracy performance and advocate fine-grained evaluations in \ac{NBR}. 

Apart from the work listed above, optimizing beyond-accuracy objectives in \ac{NBR} remains unexplored. Regarding the imbalance between repeat and exploration recommendations, \citet{li2023next} first define the repeat ratio of the recommended basket and point out the problem of repeat bias in \ac{NBR}. Furthermore, \citet{tran2024transformers} formulate the repeat bias as the deviation of the repeat ratio of the recommended baskets and ground truth baskets. We contribute by optimizing both diversity and item fairness through re-ranking, while also mitigating repeat bias to improve overall recommendation quality.

\header{Fair/diverse re-ranking.}
Re-ranking algorithms are usually designed to adjust the ranked results obtained from information access systems considering beyond-accuracy objectives. Commonly used re-ranking algorithms are constrained optimization and \ac{MMR}~\cite{carbonell1998use}. For diverse re-ranking, \citet{zhang2008avoiding} summarize three classic patterns for improving the diversity of recommendation lists as constrained optimization problems: 
\begin{enumerate*}[label=(\roman*)]
    \item maximize the diversity under the constraint of relevance tolerance;
    \item maximize relevance under the constraint of diversity tolerance; and
    \item maximize the weighted sum of relevance and diversity.
\end{enumerate*}
In terms of fair re-ranking, \citet{biega2018equity} propose an online optimization approach that uses \ac{ILP} for re-ranking based on accumulated attention and relevance scores while constraining according to a bound on the loss of ranking quality.
\citet{singh2018fairness} optimize a probabilistic ranking to maximize expected utility under three optional fairness constraints. 
For fair recommendation, \citet{li2021user} set the user fairness metric as a constraint to reduce the recommendation quality gap between the advantaged and disadvantaged groups. CPFair~\cite{naghiaei2022cpfair} simultaneously optimizes consumer and producer fairness in the objective function for multiple rankings.
Compared to previous work, which primarily focuses on fairness or diversity in isolation, we are the first to employ a \ac{MILP}-based re-ranking algorithm to optimize diversity or item fairness, specifically accounting for repeat bias in \acs{NBR} task.

\negskip

\section{Method}
\label{setting}

Our notation is summarized in Table~\ref{tab:notations}.

\subsection{Problem formulation}
\label{formulation}
In \ac{NBR}, given a user set $U$ and an item set $I$, for each user $u \in U$, the purchase history is represented as a sequence of baskets $[ B^1_u, B^2_u, \dots, B^t_u]$, where each basket contains a set of items $B^t_u = \{i_1, i_2, \ldots, i_m|i \in I\}$. 
In \ac{NBR} the task is to predict the next basket $B^{t+1}_u$ for each user. Following the common setting, the size of the recommended basket is fixed as $K$. Generally, \ac{NBR} methods generate an item ranking $L(u)$ for each user based on the predicted relevance score $S_{ui}$, and then select the top-$K$ relevant items to form the next basket, i.e., $B^{t+1}_u = L_K(u)$.

\subsection{Model overview}
In the proposed re-ranking procedure, we select top-$N$ relevant items $L_N(u)$ for each user as candidates. The relevance scores of candidate items of all users can be represented as $S = [S_{ui}]_{|U|\times N}$. A new basket $L'_K(u)$ is selected from the candidate list $L_N(u)$ taking into account the predicted relevance, as well as the diversity, item fairness, and repeat bias of the basket. $L'_K(u) \subset L_N(u), N > K$. 
The re-ranking procedure can be formalized as a \ac{MILP} problem, which is a mathematical optimization and can be solved by heuristic algorithms and optimization solvers. In our \ac{MILP} formulation, the objective function is designed to maximize relevance, diversity (or item fairness) and reduce repeat bias simultaneously. $W$ is a binary decision matrix, $W = [W_{ui}]_{|U|\times N}$. The element $W_{ui} = 1$ indicates recommending item $i$ to user $u$, and 0 otherwise. The proposed \ac{RADiv} and \ac{RAIF} algorithms are specifically adapted for different families of \ac{NBR} methods: unified methods and combined methods. 

\subsection{Optimization objectives}
\label{objective}
In this section, we introduce the definition of three optimization objectives that we aim to achieve.

\header{Diversity.} 
Diverse recommendation aims to recommend items of various and different categories to users. In this work, we use the \ac{DS}~\cite{liang2021enhancing} as the diversity objective, computed by dividing the number of recommended categories by the basket size $K$ given the decision matrix $W$. Diversity objective $\operatorname{DS}(C, W)$ is the sum of \ac{DS} among all users:
\begin{equation}
\textstyle
    \operatorname{DS}(C, W) = \sum_{u \in U} \frac{(\# \text{recommended categories}\mid W)}{K}.
\end{equation}

\header{Item fairness.}
We evaluate the item fairness between popular item group $I_1$ and unpopular group $I_2$. The recommended items will receive exposure related to the position. \Ac{DP}~\cite{singh2018fairness} defines fairness of exposure as equal average exposure between the two groups. The average exposure of a group $I_k$ is computed as: 
\begin{equation}
\textstyle
    Exposure(I_k\mid W) = \frac{1}{|I_k|}\sum_{i \in I_k} Exposure(i\mid W).
\end{equation}
Inspired by the construction in~\cite{naghiaei2022cpfair}, the item fairness objective is designed as the difference of average exposure between group $I_1$ and $I_2$. The closer the $IF$ value is to zero, the fairer a ranking is.

\begin{equation}
    \operatorname{IF}(I_1, I_2, W) = Exposure(I_1\mid W) - Exposure(I_2\mid W).
\end{equation}

\header{Repeat bias.}
Some \ac{NBR} methods are either biased to recommending too many repeat items or explore items compared with the ground truth baskets. We aim to mitigate the repeat bias of recommended baskets. Since the repeat ratio of ground truth baskets $RepRatio_{gt}$ is unknown, we directly choose the repeat ratio of recommended baskets $RepRatio_{rec}$ as the optimization objective. Each user has a repeat item set $I^{rep}_{u}$, which contains all the items the user has bought before. Given the decision matrix $W$, the repeat ratio objective can be expressed as:
\begin{equation}
\textstyle
    RepRatio_{rec}(I^{rep}_{u},W) = \sum_{u \in U}\frac{(|L'_K(u) \cap I^{rep}_{u}| \,\mid W)}{K}~.
\end{equation}

\begin{table}[t]
\vspace{-2mm}
\caption{Notation used in the paper.
}
\label{tab:notations}
\centering
\begin{tabular}{@{}ll ll@{} }
\toprule
$u\in U$ & Users &
$i\in I$ & Items     \\
$RepRatio_{rec}$ & Repeat ratio of recommendation   &
$K$ & Basket size  \\
$RepRatio_{gt}$ & Repeat ratio of ground truth   &
$N$ &Number of item candidates     \\ 
$W^r$ & Decision matrix of repeat items   &
$C$ &Categories of items\\
 $W^e$ & Decision matrix of explore items &
$I^{rep}_u$ & Repeat item set of user $u$\\
$S$ & Predicted relevance matrix &$W$ & Decision matrix \\
$S^r$ & Predicted relevance matrix of repeat items &$I_1$ & Popular item group \\
$S^e$ & Predicted relevance matrix of explore items &$I_2$ & Unpopular item group \\
$L'_K(u)$& New basket of user $u$ after re-ranking &$L_N(u)$ & Item candidate list of user $u$  \\

$H(\theta)$  &Number of repeat items in combined baskets  \\

\bottomrule
\end{tabular}
\end{table}

\subsection{Repeat-bias-aware optimization for unified \ac{NBR} methods}
\label{unified}
In this section, we post-process the recommendation lists of users generated by unified \ac{NBR} methods, where a unified model generates the relevance scores of all items. We select top-$N$ relevant items of each user as candidates. The relevance scores of these item candidates can be represented as $S = [S_{ui}]_{|U|\times N}$. We apply a \ac{MILP} model to re-rank the top $N$ candidates for each user. \ac{RADiv} algorithm simultaneously optimizes diversity and repeat ratio as shown in Eq.~\ref{direct_diversity}: 
\begin{align}
\begin{split}
&\max  \frac{1}{K} \sum_{u\in U}\sum_{i=1}^{N} S_{ui}W_{ui} + \epsilon_1 \operatorname{DS}(C, W) - \lambda RepRatio_{rec}(I^{rep}_{u},W) \\
&\text{ such that } \sum_{i=1}^{N} W_{ui} = K, W_{ui} \in \{0, 1\}.
\label{direct_diversity}
\end{split}
\end{align}
We design the objective function to maximize the sum of relevance scores, the diversity score $\operatorname{DS}(C, W)$, while minimizing the $RepRatio_{rec}(I^{rep}_{u}, W)$. Here, $\epsilon_1$ and $\lambda$ are the weighting parameters of the diversity term and repeat ratio term, respectively. $W$ is a binary decision matrix determining whether to recommend item $i$ to user $u$. The constraint indicates that the algorithm ultimately recommends $K$ items to each user, which is the basket size. It is worth noting that the minus sign in front of $RepRatio_{rec}$ is designed for repeat-biased \ac{NBR} methods with quite high $RepRatio_{rec}$. The minus implies that the algorithm aims to reduce the $RepRatio_{rec}$ to be closer to $RepRatio_{gt}$.
In contrast, for explore-biased methods with quite low $RepRatio_{rec}$ values, this term should be set to a positive sign: $+ \lambda RepRatio_{rec}(I^{rep}_{u},W)$. The algorithm will increase the $RepRatio_{rec}$ so as to approach $RepRatio_{gt}$. 

Similarly, \ac{RAIF} simultaneously optimizes item fairness and repeat ratio by Eq.~\ref{direct_fair}. 
\begin{align}
\begin{split}
& \max \sum_{u\in U}\sum_{i=1}^{N} S_{ui}W_{ui} - \alpha_1 \operatorname{IF}(I_1, I_2, W) - \lambda RepRatio_{rec}(I^{rep}_{u},W) \\
&\text{ such that } \sum_{i=1}^{N} W_{ui} = K, W_{ui} \in \{0, 1\}.
\label{direct_fair}
\end{split}
\end{align} 
The objective function is designed to maximize the combination of relevance scores, item fairness $\operatorname{IF}(I_1, I_2, W)$, and minimize $RepRatio_{rec}(I^{rep}_{u},W)$. Here, $\alpha_1$ and $\lambda$ are the weighting parameters of item fairness term and repeat ratio term, respectively. 

\subsection{Repeat-bias-aware optimization for combined \ac{NBR} methods}
\label{combination}
In the previous setting, the optimization algorithm is designed for unified \ac{NBR} methods, where the relevance scores of repeat items and explore items are generated by the same model and are comparable. However, there is another \ac{NBR} paradigm, such as \ac{TREx}~\cite{li2024we}, where the repeat item list and explore item list are obtained from different models. 
This paradigm allows one to combine the strongest repetition model and the strongest exploration model flexibly, and adjust the proportion of these two parts.
When we optimize diversity (or item fairness) and repeat bias of combined \ac{NBR} methods, the challenge lies in the fact that the predicted relevance scores are not comparable between repeat items and explore items. 

Inspired by the TREx framework \cite{li2024we}, we use a threshold $\theta$ to filter the repeat item candidates and count the number of repeat items with a relevance score larger than $\theta$. The numbers of repeat items for each user are saved as vector $H(\theta)_{\text{aux}}$. The final repeat position number for each user is $H(\theta) = \min (H(\theta)_{\text{aux}}, K)$. Here, $\theta$ indirectly controls $RepRatio_{rec}(\theta)$. The higher the $\theta$ becomes, the lower the number of repeat slots $H(\theta)$ and $RepRatio_{rec}(\theta)$ become. $H(\theta)$ is an important variable to avoid the comparison between repeat and explore relevance scores. In this way, repeat items are compared internally and $H(\theta)$ of them are finally selected. Explore items compete internally and $K-H(\theta)$ of them are chosen.

The \ac{RADiv} algorithm is adapted for combined \ac{NBR} methods as shown in Eq.~\ref{combine_div}. $S^r = [S^r_{ui}]_{|U|\times N}$ and $S^e = [S^e_{ui}]_{|U|\times N}$ are predicted relevance score matrices of repeat items and explore items from different recommender systems, respectively. $W^r = [W^r_{ui}]_{|U|\times N}$, $W^e = [W^e_{ui}]_{|U|\times N}$ are corresponding binary decision matrices to determine the selection of repeat items and explore items. $\epsilon_2$ is the weighting parameter of the diversity term. The constraints indicate the repeat slots for each user are $H(\theta)$, while the explore slots for each user are $K - H(\theta)$. Similarly, \ac{RAIF} is adjusted to optimize item fairness $\operatorname{IF}(I_1, I_2, W^r, W^e)$ and $RepRatio_{rec}(\theta)$ in Eq.~\ref{combine_fair}. $\alpha_2$ is the weighting parameter of item fairness term. Algorithm \ref{ilp} summarizes our algorithms for combined \ac{NBR} methods.
\begin{align}
\begin{split}
&\max \frac{1}{K} \sum_{u\in U}\sum_{i=1}^{N} (S_{ui}^{r}W_{ui}^{r} + S_{ui}^{e}W_{ui}^{e}) + \epsilon_2 \operatorname{DS}(C, W^r, W^e) \\
&\mbox{}\hspace*{5mm}\text{ such that } \sum_{i=1}^{N} W_{ui}^{r} = H(\theta), \sum_{i=1}^{N} W_{ui}^{e} = K - H(\theta), W_{ui}^{r}, W_{ui}^{e} \in \{0, 1\};
\label{combine_div}
\end{split}
\\
\begin{split}
& \max \sum_{u\in U}\sum_{i=1}^{N} (S_{ui}^{r}W_{ui}^{r} + S_{ui}^{e}W_{ui}^{e}) - \alpha_2 \operatorname{IF}(I_1, I_2, W^r, W^e) \\
&\mbox{}\hspace*{5mm}\text{ such that } \sum_{i=1}^{N} W_{ui}^{r} = H(\theta), \sum_{i=1}^{N} W_{ui}^{e} = K - H(\theta), W_{ui}^{r}, W_{ui}^{e} \in \{0, 1\}.
\label{combine_fair}
\end{split}
\end{align}

\begin{algorithm}[!t]
	\SetAlgoLined
	\KwIn{Basket size $K$, Categories $C$, Number of item candidates $N$, Item group $I_1,I_2$, 
        Parameters $\epsilon_2, \alpha_2, \theta$}
	\KwOut{Recommendation matrix $W^r, W^e$}
	
	$S^r, S^e \leftarrow$ The top-$N$ repeat and explore relevance scores of users.  \label{alg:line1}

      Compare $\theta$ with $S^r_{ui}$, and obtain $H(\theta) = \min (H(\theta)_{\text{aux}}, K)$.
 
      Solve the optimization problem following Eq. \ref{combine_div} (or \ref{combine_fair}).

\textbf{Return} $W^r, W^e$
	\caption{\ac{RADiv} and \ac{RAIF} for Combined \ac{NBR} Methods}
	\label{ilp}
\end{algorithm}

\vspace*{-2mm}
\section{Experimental Setup}
\header{\ac{NBR} methods.}
We select and investigate the following 5 representative \acs{NBR} methods:

\begin{itemize}[leftmargin=*,nosep] 

\item \textbf{UP-CF@r}, which combines recency-aware user-wise popularity and collaborative filtering while considering the recent shopping behavior~\citep{faggioli2020recency}.
\item \textbf{TIFUKNN}, which models the temporal dynamics of users' past baskets by using a KNN-based approach based on the \ac{PIF}~\citep{hu2020modeling}. 
\item \textbf{Dream}, which forms basket representations using a pooling strategy and models sequential user behavior through an RNN~\citep{yu2016dynamic}. 
\item \textbf{DNNTSP} uses \ac{GNN} and self-attention mechanisms to encode item-to-item relations across baskets and capture temporal dependencies~\citep{yu2020predicting}. 
\item \textbf{\ac{TREx}}, which has a repetition module considering item repurchase features and users' interests, and an exploration module targeted for beyond-accuracy metrics. Then, repeat items and explore items are combined to form the final basket~\cite{li2024we}.
\end{itemize}

\noindent%
We exclude \acs{NBR} methods that only focus on the repetition recommendation (i.e., P-TopFreq, ReCANet~\citep{ariannezhad2022recanet}, and NBRR~\citep{katz2022learning}) or exploration recommendation (i.e., NNBR \citep{li2023masked}), as we aim to investigate and balance repetition \emph{and} exploration in \acs{NBR}. 

\header{Datasets.}
Following previous \acs{NBR} studies~\citep{liu2024measuring, li2024we, li2023masked, li2023next}, we select three publicly available grocery shopping datasets: TaFeng~\citep{tafeng-2001-dataset}, Dunnhumby~\citep{dunnhumby-2024-source}, and Instacart~\citep{instacart-2017-dataset}. 
They exhibit different characteristics in terms of repetition and exploration, which are critical for verifying the effectiveness of our proposed repeat-bias-aware optimization algorithms.

For each dataset, we remove users with fewer than three baskets and items purchased fewer than five times, as done in \cite{ariannezhad2022recanet}. Due to the large size of the Instacart dataset, memory limitations occurred during the calculation of some methods. Following~\cite{naumov2023time}, we also randomly sampled 20,000 users from Instacart before applying any filtering. Table~\ref{dataset} provides the statistics of the three datasets after preprocessing. The average $RepRatio_{gt}$ refers to the average proportion of repeat items in the ground truth baskets, as defined in \cite{li2023next}.

We split each dataset following the approach in \cite{ariannezhad2022recanet,faggioli2020recency,naumov2023time}. The training set includes all baskets from each user except the last basket. For users with more than 50 baskets in the training data, we limit the training set to their most recent 50 baskets~\cite{li2023next}. The last baskets of all users are then randomly split equally into a validation set (50\%) and a test set (50\%). \looseness=-1

\begin{table}[t]
\centering
\setlength{\tabcolsep}{2.8pt}
\caption{Statistics of the datasets after preprocessing.}
\label{dataset}
\begin{tabular}{l cccccc}
\toprule
& & & & Avg.  & Avg. & Avg. \\
Dataset &  \#Users & \#Items & \#Baskets  & \#baskets/user  & \#items/basket & $RepRatio_{gt}$\\  
\midrule
Instacart & 19,210 &  29,399 & 305,582 & 15.91  & 10.06 & 0.60 \\
Dunnhumby & \phantom{0}2,482  & 37,162 & 107,152 & 43.17  & 10.07   & 0.43 \\
TaFeng    &10,182 & 15,024 & \phantom{0}82,387 & \phantom{0}8.09  & \phantom{0}6.14 & 0.21 \\  
\bottomrule
\end{tabular}
\end{table}

\header{Evaluation metrics.}
We use the following widely used metrics in our experiments to measure how the proposed algorithms balance multiple objectives. In terms of accuracy, Recall measures the system's ability to retrieve items that users will purchase in their next baskets. For item fairness, demographic parity (DP) measures the ratio of the average exposure of the popular group to the average exposure of the unpopular group. Following~\cite{raj2022measuring}, we use logDP to deal with the empty-group case. The closer the logDP value is to zero, the fairer the recommendation is. The diversity score (DS) measures the number of categories within the recommended basket divided by basket size. Repeat bias is computed as: RepBias = $RepRatio_{rec}-RepRatio_{gt}$~\cite{tran2024transformers}. Following~\cite{naghiaei2022cpfair}, we use a comprehensive metric to evaluate the overall performance of item fairness and RepBias: mFR = $\omega |\text{logDP}| + (1-\omega)|\text{RepBias}|$. Similarly, the overall performance of diversity and RepBias is evaluated as $\text{mDR} = \omega \text{DS} - (1-\omega)|\text{RepBias}|$.

\header{Implementation details.}
In our experiments, we follow~\cite{hu2020modeling,li2024we} and set basket size $K=20$. The number of item candidates $N =100$. We compute the popularity of all items and select the top 20\% as popular $I_1$ and the rest belong to the unpopular group $I_2$. We apply grid search to select hyperparameters on the validation set. We select $\alpha_1$ and $\alpha_2$ in [0, 0.001, 0.01, 0.1, 1, 10, 20, 30, 40, 50, 60, 70, 80, 90, 100, 200]. $\epsilon_1$ and select $\epsilon_2$ from [0, 0.001, 0.01, 0.02, 0.04, 0.06, 0.08, 0.1, 0.12, 0.14, 0.16, 0.18, 0.2]. We choose $\lambda$ in [0, 0.001, 0.01, 0.1, 0.2, 0.3, 0.4, 0.5, 0.6, 0.7, 0.8, 0.9, 1]. $\theta$ candidates are selected based on $S^r$. 

We select the optimal hyperparameter combination according to these two conditions: 
\begin{enumerate*}[label=(\roman*)]
    \item satisfy a 10\% Recall drop tolerance; and
    \item maximize mDR or minimize mFR, where $\omega = 0.5$ following~\cite{naghiaei2022cpfair}.
\end{enumerate*}
We use Gurobi,\footnote{\url{https://www.gurobi.com}} which is an industrial optimization solver capable of delivering practical and effective feasible solutions. We release our code and hyperparameters at \url{https://github.com/lynEcho/Repbias_NBR}.

\section{Experimental Results}
\label{results}

We design experiments to answer three research questions: 
\begin{enumerate*}[label=\textbf{RQ\arabic*},nosep,leftmargin=*]
    \item How does optimizing only for diversity or item fairness affect the repeat bias? \label{rq1} 
    
    \item What is the performance of proposed \ac{RADiv} and \ac{RAIF} algorithms on \ac{NBR} methods? \label{rq2}
    
    \item How does \ac{RADiv} and \ac{RAIF} algorithms strike a balance between utility and beyond-accuracy metrics? \label{rq3}   
\end{enumerate*}

\begin{figure}[t]
    \centering
    
    \begin{subfigure}{\textwidth}
        \centering
        \includegraphics[width=0.9\linewidth]{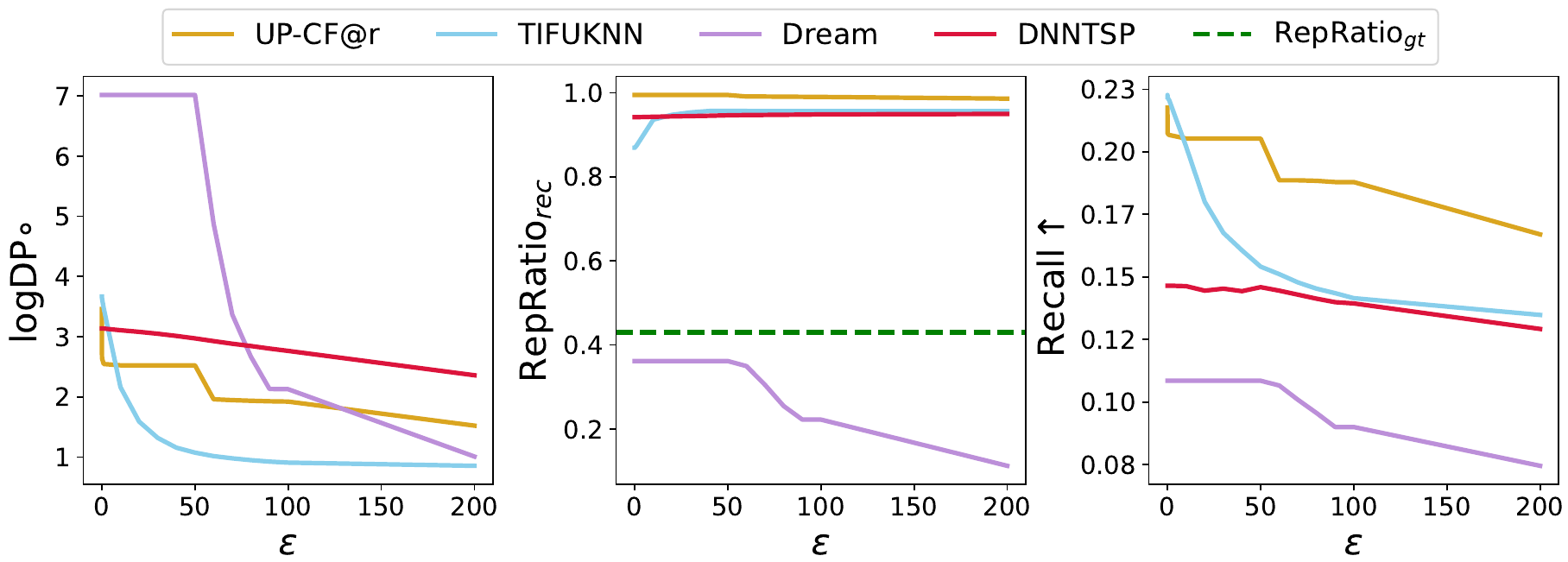} 
        \vspace*{-2mm}
        \caption{Item fairness optimization}
        \label{naive_fair}
    \end{subfigure}
    
    \vspace{1mm}
    
    \begin{subfigure}{\textwidth}
        \centering
        \includegraphics[width=0.9\linewidth]{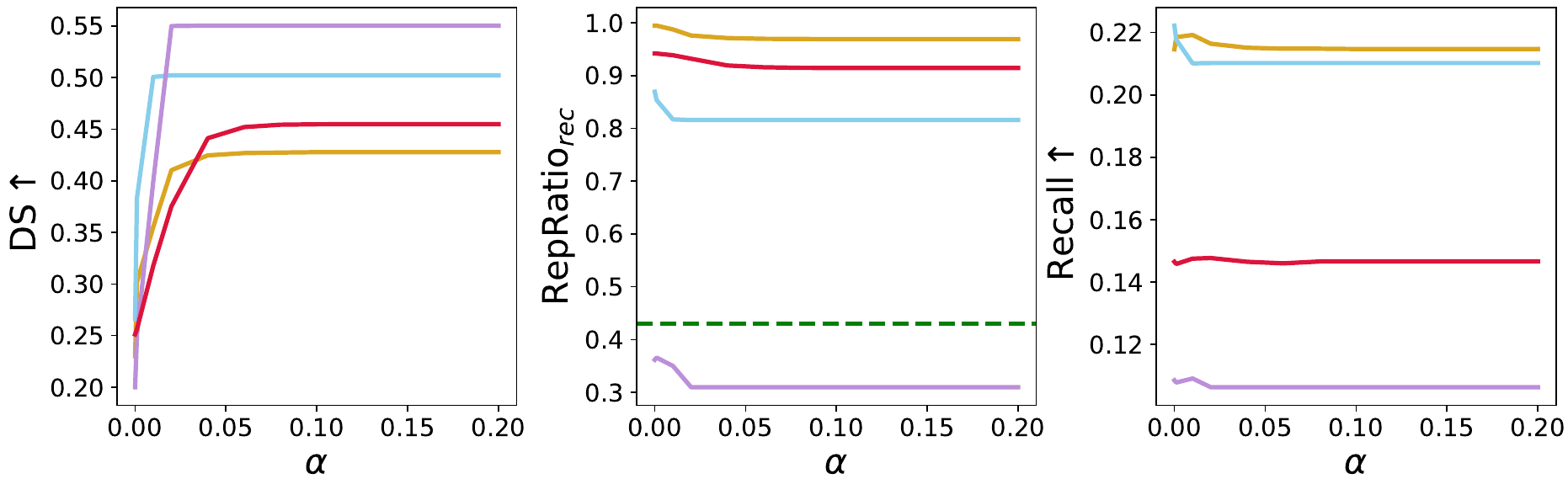}  
        \vspace*{-2mm}
        \caption{Diversity optimization}
        \label{naive_div}
    \end{subfigure}

    \vspace{1mm}
    
    \caption{Performance of item fairness and diversity optimization without considering $RepRatio_{rec}$ on different NBR methods. logDP$\circ$ means the closer logDP value is to zero, the fairer it is.} 
    \label{naive}
\end{figure}

We start by studying how optimizing for beyond-accuracy metrics, without taking the repeat bias, would affect the baskets in terms of repeat-explore items, answering \ref{rq1}.
We apply a re-ranking algorithm without considering $RepRatio_{rec}$ to optimize the recommended baskets obtained from different \ac{NBR} methods on Instacart, Dunnhumby, and TaFeng. For diverse re-ranking, the optimization objective is formulated as $\max  \frac{1}{K} \sum_{u\in U}\sum_{i=1}^{N} S_{ui}W_{ui} + \epsilon \operatorname{DS}(C, W)$. In terms of fair re-ranking, the objective function is $\max \sum_{u\in U}\sum_{i=1}^{N} S_{ui}W_{ui} - \alpha \operatorname{IF}(I_1, I_2, W)$. The parameters $\epsilon$ and $\alpha$ are used to adjust the weight of the diversity and item fairness terms, respectively. Fig.~\ref{naive} shows the performance of diversity and item fairness optimization on Dunnhumby.\footnote{We observe a similar trend on the Instacart and TaFeng datasets. Because of space limitations, we report the results on Instacart and TaFeng in the anonymous repository.} \looseness=-1

We make the following observation:
\begin{enumerate*}[label=(\roman*)]
    \item Diversity and item fairness optimizations significantly improve the \ac{NBR} methods in terms of logDP and DS.
    For item fairness optimization, as the parameter $\epsilon$ increases ($0\rightarrow 200$), all \ac{NBR} methods become fairer. Also, the \ac{DS} of all \ac{NBR} methods increase as $\alpha$ increases ($0 \rightarrow 0.2$).
    
    \item In most cases, Recall exhibits a declining trend as DS and logDP become better. 
    In diversity optimization, the Recall of UP-CF@r, Dream, and DNNTSP improve slightly when $\alpha = 0.01$.

    \item In terms of repeat bias, which is measured as the gap between $RepRatio_{rec}$ and $RepRatio_{gt}$ (green dashed line in Fig.~\ref{naive}). We see that the repeat bias is likely to either intensify or diminish, showing no clear trend. In item fairness optimization, the repeat bias of TIFUKNN and Dream grows as $\epsilon$ increases, showing that it is important to consider repeat bias while optimizing beyond-accuracy metrics. 
\end{enumerate*}

To answer \ref{rq2}, we apply \ac{RADiv} and \ac{RAIF} algorithms on five \ac{NBR} methods. Of these, TIFUKNN, UP-CF@r, DNNTSP, and Dream are unified repeat-explore recommendation methods while \ac{TREx} is a combined method. 
Since \ac{TREx} does not provide predicted relevance scores for explore items, here we use the explore item list obtained from UP-CF@r to combine with repeat items of \ac{TREx}. 
Table~\ref{instacart} reports the results on Instacart.\footnote{Similar patterns are observed on the Dunnhumby and TaFeng datasets. Due to space limitations, we report the results on Dunnhumby and TaFeng in the anonymous repository.} We also conduct an ablation study, i.e., optimizing only for diversity, item fairness, and repeat bias, respectively. 

We arrive at the following conclusions:
\begin{enumerate*}[label=(\roman*)]
    \item The proposed \ac{RADiv} algorithm (RD) achieves a significant improvement in DS, RepBias, and comprehensive metric mDR in comparison to the original (Ori.) results on all \ac{NBR} methods. Surprisingly, the Recall of Dream increases after \ac{RADiv} optimization ($0.0977 \rightarrow 0.1468$).
  
    \item The \ac{RAIF} algorithm (RF) achieves better logDP, less RepBias, and better overall metric mFR than the original (Ori.) results, proving the effectiveness of \ac{RAIF} algorithm.

    \item \ac{RADiv} and \ac{RAIF} algorithms can effectively adjust the $RepRatio_{rec}$ of methods to approach $RepRatio_{gt}$. Take Instacart as an example, \ac{RADiv} and \ac{RAIF} algorithms decrease the $RepRatio_{rec}$ of TIFUKNN, UP-CF@r, DNNTSP, and increase the $RepRatio_{rec}$ of Dream to make them closer to ground truth 0.6.

    \item In our ablation study, we find that only optimizing $RepRatio_{rec}$ (R) will decrease the diversity and item fairness of TIFUKNN, UC-CF@r, DNNTSP and \ac{TREx}. This means that it is necessary to optimize both $RepRatio_{rec}$ and diversity, both $RepRatio_{rec}$ and item fairness, as implemented by the proposed \ac{RADiv} and \ac{RAIF} algorithms.
\end{enumerate*}
\begin{table}[t!]
\caption{The performance of \ac{RADiv} and \ac{RAIF} algorithms on Instacart ($RepRatio_{gt} = 0.6$). \textit{Ori.} indicates the original baskets obtained from each method. \textit{RD} indicates \ac{RADiv}. \textit{RF} refers to \ac{RAIF}. In ablation study, \textit{D}, \textit{F}, \textit{R} indicate optimizing only for diversity, item fairness, and repeat bias, respectively. RepR indicates $RepRatio_{rec}$ (best close to $RepRatio_{gt}$). RepBias and logDP (best close to zero).}
\label{instacart}
\resizebox{\linewidth}{!}{\begin{tabular}{lllccrrlllccl@{}}
\toprule
\multirow{2}{*}{\rotatebox[origin=c]{90}{Meth.}}     & \multicolumn{6}{c}{Diversity optimization}  & \multicolumn{6}{c}{Item fairness optimization} 
\\ 
\cmidrule(r){2-7} \cmidrule(r){8-13}
& Type & Recall$\uparrow$  & DS$\uparrow$     & RepR & RepBias &mDR$\uparrow$ &Type  & Recall$\uparrow$  & logDP   & RepR & RepBias &mFR$\downarrow$
\\ \midrule
\multirow{4}{*}{\rotatebox[origin=c]{90}{TIFUKNN }}    &  Ori.    & \textbf{0.4559}       & 0.3615     & 0.9248        &0.3248 &  0.0184       &  Ori.     & \textbf{0.4559}     &   3.1252     &    0.9248    & 0.3248   &  1.7250    \\
                      & D     &  0.4259     & 0.5897     & 0.8984       &  0.2984    &  0.1457  & F      &  0.4269      & 2.3510        &   0.9252     & 0.3252    & 1.3381    \\ 
                      &   R   &  0.4537   &0.3587    &0.9100       &  0.3100  &  0.0244    &   R    &  0.4232      &  3.2760     & \textbf{0.7939}   &  \textbf{0.1939}  & 1.7350  \\
                      &   RD   &   0.4245   & \textbf{0.5898}   &\textbf{0.8874}         & \textbf{0.2874} &  \textbf{0.1512}  &   RF    & 0.4098     &\textbf{2.3271}   & 0.8718    & 0.2718     & \textbf{1.2995} \\ \midrule
\multirow{4}{*}{\rotatebox[origin=c]{90}{UP-CF@r}} & Ori.    & \textbf{0.4405} & 0.3489 &0.8905  & 0.2905& 0.0292  & Ori.    & \textbf{0.4405} & 3.3966 &0.8905   &0.2905 &1.8436  \\
                      & D    & 0.3983    & \textbf{0.6375}       &0.7896          &  0.1896     & 0.2239  & F     &  0.4282      & 2.4860       &   0.8858       & 0.2858     &  1.3859  \\
                      & R    & 0.4353      &0.3424      &0.8812          &  0.2812    &  0.0306  & R     &  0.4373      &  3.3997       &    0.8791     &0.2791      & 1.8394   \\
                      & RD   &  0.3968     & \textbf{0.6375}      &\textbf{0.7837}   & \textbf{0.1837}     & \textbf{0.2269}  & RF   & 0.4052    &  \textbf{2.1810}     & \textbf{0.8019}    &  \textbf{0.2019}  &\textbf{1.1915}     \\ \midrule
\multirow{4}{*}{\rotatebox[origin=c]{90}{DNNTSP}}    &  Ori.    & \textbf{0.4347}   & 0.3402   & 0.9133     & 0.3133    & 0.0135    &  Ori.    & \textbf{0.4347}       &   3.2573    & 0.9133     & 0.3133  &  1.7853     \\
                      & D     & 0.4046      & 0.6009      & 0.8660        &  0.2660   &  0.1675   &  F     & 0.4332     & \textbf{2.9780}     &  0.9152    &  0.3152   &  1.6466   \\
                      &  R    & 0.4337  & 0.3388  &  0.9080       &0.3080     &  0.0154   &   R    &  0.4256     &    3.3204    &  \textbf{0.8640}    &  \textbf{0.2640}    &  1.7922 \\
                      &  RD    &  0.4059   & \textbf{0.6013}    &\textbf{0.8623}        & \textbf{0.2623}   &  \textbf{0.1695}   & RF     &  0.4277      & 2.9901    & 0.8837  &  0.2837    & \textbf{1.6369} \\ \midrule
\multirow{4}{*}{\rotatebox[origin=c]{90}{Dream}}    &  Ori.    &  0.0977    &0.1000    & 0.1923      &   -0.4077   &  -0.1539  &  Ori.    &  0.0977     &  7.3111     &  0.1923        &  -0.4077      &  3.8594\\
                      & D     & 0.0723     &\textbf{0.7000}    & 0.1267        &  -0.4733    & 0.1133   &   F    &   0.0709     &\textbf{2.1018}  &  0.1226    &  -0.4774      & \textbf{1.2896}\\
                      & R     &\textbf{0.1540}   & 0.1243     & \textbf{0.4202}        & \textbf{-0.1798}    & -0.0277    &   R    & \textbf{0.1288}     & 7.3111     & \textbf{0.3108}    & \textbf{-0.2892}   & 3.8002     \\
                      &  RD  & 0.1468  &  0.5654    & \textbf{0.4202}        & \textbf{-0.1798}    &\textbf{0.1928}    &  RF     & 0.0939   & 2.3893    &  0.2094     &  -0.3906     &  1.3900 \\ \midrule

\multirow{4}{*}{\rotatebox[origin=c]{90}{TREx}}    &Ori.     &  \textbf{0.4595}      & 0.3533       &  0.9265        &  0.3265   &  0.0134   &   Ori.   & \textbf{0.4595}       &  3.2182       &  0.9265        &   0.3265    & 1.7724  \\
                      &  D    &   0.4455     & 0.5296       &  0.9265        &  0.3265    & 0.1016   &  F     &  0.4569       & 2.9052        &  0.9265        &  0.3265     & 1.6159  \\
                      &   R   &  0.4368      & 0.3280       & \textbf{0.7973}        & \textbf{0.1973}   &  0.0654   &   R    &  0.4277       & 3.4281        &\textbf{0.7487}         & \textbf{0.1487}      & 1.7884  \\
                      &  RD    &   0.4195     & \textbf{0.5874}       &   \textbf{0.7973}      & \textbf{0.1973}    & \textbf{0.1951}    &  RF     &   0.4226      & \textbf{2.6528}       & \textbf{0.7487}         & \textbf{0.1487}    & \textbf{1.4008 }   \\ \bottomrule
\end{tabular}}
\end{table}

\begin{figure}[t]
    \centering
    
    \begin{subfigure}{\textwidth}
        \centering
        \includegraphics[width=1\linewidth]{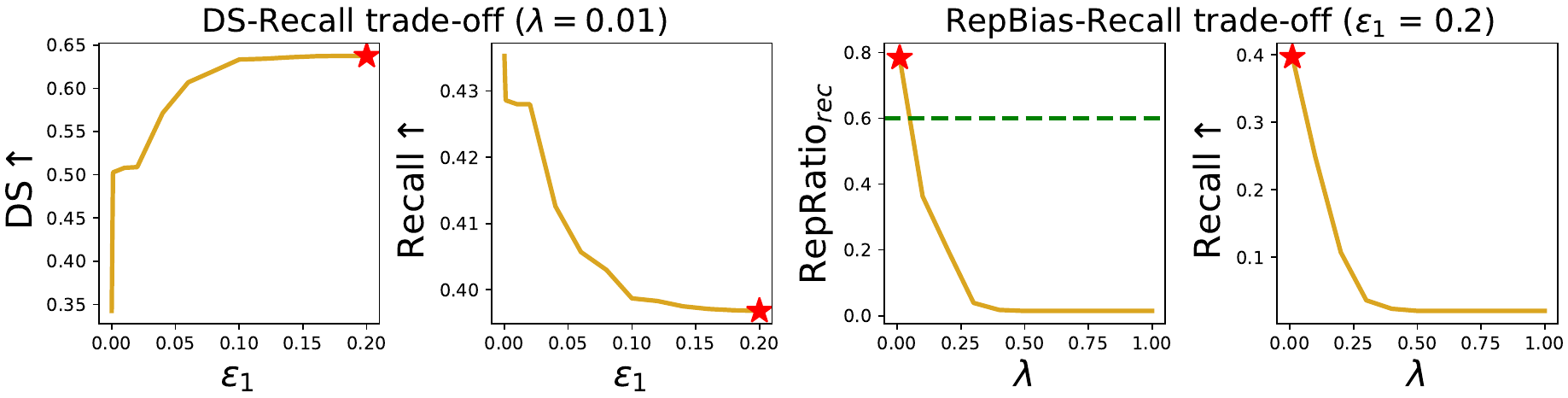} 
        \vspace*{-6mm}
        \caption{\ac{RADiv} optimization}
        \label{trade_ds}
    \end{subfigure}
    
    \vspace*{1mm}
    
    \begin{subfigure}{\textwidth}
        \centering
        \includegraphics[width=1\linewidth]{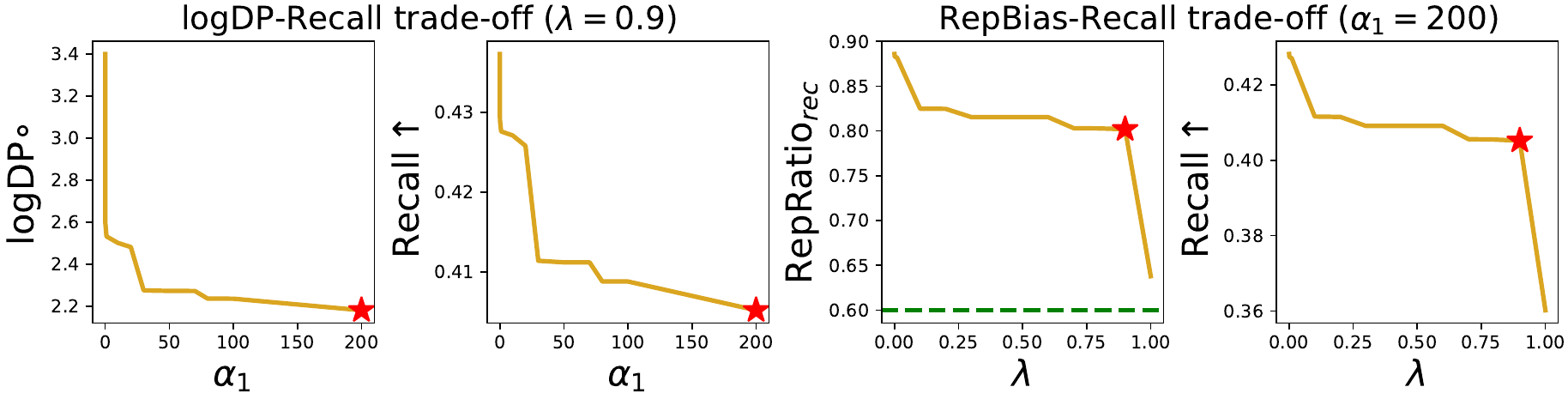}  
        \vspace*{-6mm}
        \caption{\ac{RAIF} optimization}
        \label{trade_fair}
    \end{subfigure}

    \vspace*{1mm}
    
    \caption{Trade-off between Recall and diversity, item fairness, repeat bias. Take UP-CF@r as an example on the Instacart dataset. The red star indicates the optimal solution chosen by the proposed algorithm. The green dashed line is value of $RepRatio_{gt}$.}
    \label{trade}
\end{figure}

To answer \ref{rq3}, we take UP-CF@r as an example to illustrate the trade-off between Recall and other metrics on Instacart. For the two parameters $\epsilon_1$ and $\lambda$ in the \ac{RADiv} optimization, we change one parameter each time with the other one fixed. Similarly, we perform the same operation for $\alpha_1$ and $\lambda$ in \ac{RAIF} optimization. The analysis is shown in Fig.~\ref{trade}.

We make the following observations: 
\begin{enumerate*}[label=(\roman*)]
    \item In Fig.~\ref{trade_ds}, we first fix $\lambda$ and change $\epsilon_1$ from 0 to 0.2. We see that DS continuously increases at the cost of Recall. The red star indicates that the algorithm chooses $\epsilon_1 = 0.2$ as the optimal solution to achieve a balance between diversity and Recall. When we fix $\epsilon_1$ and change $\lambda$ from 0 to 1, $RepRatio_{rec}$ and Recall decline consistently. Since our algorithm optimizes $RepRatio_{rec}$ to be approaching $RepRatio_{gt}$, it selects $\lambda=0.01$ as the optimal trade-off between Recall and repeat bias. This process reflects the balanced strategy of \ac{RADiv} to maximize diversity and mitigate repeat bias within the tolerance of utility loss. 

    \item In Fig.~\ref{trade_fair}, we fix $\lambda$ and adjust $\alpha_1$ from 0 to 200. logDP decreases indicating unpopular items are assigned with more exposure. The recommendation becomes fairer at the cost of a Recall drop. The algorithm chooses $\alpha_1=200$ to achieve a balance between logDP and Recall. When we change $\lambda$ from 0 to 1 with a fixed $\alpha_1$, $RepRatio_{rec}$ and Recall show a consistent downward trend. The algorithm selects $\lambda=0.9$ to balance repeat bias and Recall. Considering the inverse relationship between Recall and logDP, repeat bias, the \ac{RAIF} algorithm decides to sacrifice a little utility in exchange for a greater return on item fairness and repeat bias.

\end{enumerate*}

\section{Conclusion} 
\label{conclusion}
\vspace*{-2mm}

We have proposed repeat-bias-aware optimization algorithms for improving diversity and item fairness while mitigating repeat bias in \ac{NBR}. Our \ac{RADiv} and \ac{RAIF} algorithms re-rank the preliminary recommended results obtained from various \ac{NBR} baselines according to different beyond-accuracy objectives and seek a balance between repeat items and explore items simultaneously while optimizing for beyond-accuracy metrics. 
We have extended our approach to multiple \ac{NBR} paradigms~\citep{li2024we}, in particular, one that fuses repeat item lists and explores item lists from separate models, allowing us to select tailored models for each list. 
We have conducted experiments on three real-world retail datasets. We find that only optimizing for diversity or item fairness will increase repeat bias in many cases, which can reduce user satisfaction. The proposed \ac{RADiv} and \ac{RAIF} algorithms can effectively optimize diversity and fairness while mitigating the repeat bias issue under an acceptable utility loss. 
Finally, we have investigated trade-offs between Recall and other beyond-accuracy metrics, including diversity, item fairness, and repeat bias.

Based on our experiments, we find that it is critical to evaluate the utility of \ac{NBR} methods and measure their repeat bias at the same time. 
In many cases, a recommendation model can achieve the best performance in terms of Recall by recommending only repeat items (i.e., highest repeat bias), but this reduces the likelihood of novelty and serendipity in recommendations, leading to filter bubble and echo chamber issues~\citep{flaxman2016filter, ge2020understanding}.
Re-ranking may be an effective solution to this problem, as we have shown that it can strike a balance between accuracy and beyond-accuracy metrics, while taking into account the repeat bias in the re-ranking process. 
However, in extreme cases where the original ranking does not include many explore items, re-ranking cannot really be effective.  
Therefore, we suggest to consider accuracy and beyond-accuracy objectives in the original ranking. In addition, the definition and quantification of repeat bias is also a direction worth exploring in the future. 

\subsection*{Acknowledgments}
\halfnegskip

This research was (partially) supported by the Dutch Research Council (NWO), under project numbers 024.004.022, NWA.1389.20.\-183, and KICH3.LTP.20.006, and the European Union's Horizon Europe program under grant agreement No 101070212.

All content represents the opinion of the authors, which is not necessarily shared or endorsed by their respective employers and/or sponsors.

\bibliographystyle{splncs04nat}
\bibliography{references}

\end{document}